\documentclass{ws-procs9x6-cpt19}

\def\al{\alpha}
\def\be{\beta}
\def\ga{\gamma}
\def\de{\delta}

\def\ve{\varepsilon}

\def\et{\eta}
\def\tht{\theta}   

\def\ka{\kappa}
\def\la{\lambda}

\def\rh{\rho}

\def\si{\sigma}

\def\ph{\phi}

\def\ps{\psi}
\def\om{\omega}
\def\Ga{\Gamma}
\def\De{\Delta}

\def\fr#1#2{{{#1} \over {#2}}}
\def\frac#1#2{{\textstyle{{#1}\over {#2}}}}

\def\prt{\partial} 

\newcommand*\lsim{\mathrel{\rlap{\lower4pt\hbox{\hskip1pt$\sim$}}
    \raise1pt\hbox{$<$}}}
\newcommand*\expect[1]{\langle{#1}\rangle}

\newcommand{\altnabla}{\widetilde{\nabla}} 

\begin{document}

\newcommand{\refeq}[1]{(\ref{#1})}
\def\etal {{\it et al.}}

\title{Noncommutative Gravity and the Standard-Model Extension}

\author{Charles D.\ Lane,$^{1,2}$}

\address{$^1$Department of Physics, Berry College,\\
Mount Berry, GA 30149, U.S.A.}

\address{$^2$IU Center for Spacetime Symmetries,\\
Bloomington, IN 47405, U.S.A.}

\begin{abstract} 
Noncommutative geometry has become popular mathematics 
for describing speculative physics beyond the Standard Model. 
Noncommutative QED has long been known to fit within 
the framework of the Standard-Model Extension (SME). 
We argue in this work that noncommutative gravity 
also fits within the SME framework. 
\end{abstract}

\bodymatter

\phantom{}\vskip10pt\noindent 
The original inspiration for considering noncommutative geometry in physics\cite{ncgeom} 
was the desire to have a Heisenberg-like uncertainty relation for position coordinates: 
$\De x \De y > 0$, 
which corresponds to noncommutativity between position coordinates, 
$[x,y]\ne 0$. 
This idea may be made compatible with {\itshape observer} Lorentz symmetry 
by assuming 
$\left[ x^\mu,x^\nu \right] = i\tht^{\mu\nu}$, 
where $\tht^{\mu\nu}$ is real and antisymmetric. 
(Note that the existence of a nonzero tensor 
that appears to be a property of spacetime itself 
violates {\itshape particle} Lorentz symmetry.) 

A useful tool for constructing noncommutative theories 
is the Moyal~$\star$ product.\cite{moyalproduct} 
Consider a commutative field theory with functions/fields $f, g, \ldots$. 
This may be turned into a noncommutative field theory with 
noncommutative functions/fields $\hat{f}, \hat{g}, \ldots$ 
by replacing all ordinary products 
with $\star$ products: 
\begin{equation} 
(f\cdot g)(x) \rightarrow (\hat{f}\star\hat{g})(x) 
 := \left. \exp\left( \frac{i}{2}\tht^{\mu\nu}\frac{\prt}{\prt x^\mu}\frac{\prt}{\prt y^\nu} \right) 
      \hat{f}(x)\hat{g}(y) \right|_{x=y} 
 \quad . 
\end{equation} 
Note: (1) This automatically gives $\left[x^\mu, x^\nu\right] \rightarrow 
 \left[\widehat{x}^\mu, \widehat{x}^\nu \right]_{\star} = i\theta^{\mu\nu}$ as desired. 
(2) It has similar form to a multivariable Taylor series, 
and hence may be related to nonlocality. 
(3) The Moyal $\star$ product is not the {\itshape only} way to define a noncommutative theory; 
it is simply one convenient approach. 

Interpretation of such noncommutative theories is nontrivial as 
the noncommutative fields $\widehat{\ps},\widehat{A}_\mu,\ldots$ 
do not necessarily correspond to physical particles. 
A Seiberg-Witten map\cite{swmap} 
$\widehat{\ps},\widehat{A}_\mu,\ldots\rightarrow \ps,A_\mu,\ldots$ 
is a method of restating noncommutative gauge theories 
that eases interpretation. 
This map guarantees that $\ps,A_\mu$ are ordinary fields 
with ordinary gauge transformations 
whose behavior is physically equivalent to $\widehat{\ps},\widehat{A}_\mu$.

This strategy has been used to show that noncommutative QED\cite{bichl} 
fits within the flat-space SME.\cite{carroll} 
In the rest of this work, 
we relate a model of noncommutative gravity 
to the gravitational SME.\cite{baileylane} 

One way to model gravity is as a spontaneously broken SO(2,3) gauge theory.\cite{ciric1} 
This provides a good starting place to build a noncommutative model of gravity, 
as the (broken) gauge symmetry is automatically respected by the Seiberg-Witten map. 

The unbroken commutative SO(2,3) action on flat (1+3)-dimensional spacetime 
may be written 
$S= c_1 S_1 + c_2 S_2 + c_3 S_3$, where 
\begin{eqnarray} 
&& S_1 \sim {\rm Tr}\int d^4x\ \ve^{\mu\nu\rh\si} F_{\mu\nu} F_{\rh\si} \ph 
 , \quad 
 S_2 \sim {\rm Tr}\int d^4x\ \ve^{\mu\nu\rh\si} F_{\mu\nu} D_\rh\ph D_\si\ph\, \ph 
 , \nonumber \\ 
&& \mbox{ and } S_3 \sim {\rm Tr}\int d^4x\ \ve^{\mu\nu\rh\si}  D_\mu\ph D_\nu\ph D_\rh\ph D_\si\ph\, \ph 
 \quad . 
\end{eqnarray} 
In this expression, $F$ is the SO(2,3) gauge field, 
$D$ is the associated covariant derivative, 
$\ph$ is a scalar field, 
and $c_1,\ldots,c_3$ are undetermined weights. 

If we then assume that $\ph$ spontaneously breaks the SO(2,3) symmetry 
in its ground state, 
$\expect{\ph}=(0,0,0,0,\ell)$, 
and expand the action around this ground state, 
then it takes a form that includes conventional gravity: 
$S \supset -\fr{1}{16\pi G_N} \int d^4x\ e 
 \left( 
 R - \fr{6}{\ell^2}(1+c_2+2c_3) 
 \right) 
$. 

This model may then inspire a noncommutative gravitational theory\cite{ciric2} 
by following a similar prescription to that followed for NCQED: 
(1) Start with the unbroken SO(2,3) action. 
(2) Replace fields $F,\ph$ with Moyal~$\star$ products of noncommutative fields $\widehat{F},\widehat{\ph}$. 
(3) Apply a Seiberg-Witten map to replace noncommutative fields 
with physically equivalent commutative fields. 
(4) Assume that the SO(2,3)$_\star$ symmetry is spontaneously broken 
by $\ph$ having a nonzero vacuum expectation value. 
The resulting action is left with a noncommutative SO(1,3)$_\star$ symmetry. 
It may be expanded in powers of $\tht^{\mu\nu}$, taking the form 
\begin{equation} 
S_{\rm NCR} = 
 -\int \frac{d^4x\ e}{16\pi G_N}  
 \left\{ \left[ R - \frac{6(1+c_2+2c_3)}{\ell^2} \right]  
 + \frac{1}{8\ell^4} \sum_{u=1}^{6} 
  \theta^{\al\be} \theta^{\ga\de} C_{(u)} L^{(u)}_{\al\be\ga\de} 
 \right\} 
 \quad . 
 \label{NCaction} 
\end{equation} 
The initial bracketed term describes conventional General Relativity. 
The noncommutative modification is a sum of geometric quantities $L^{(u)}$ 
and their weights $C_{(u)}$, 
which are listed in Table \ref{table1}. 
\begin{table} 
\tbl{Quantities appearing in the noncommutative action.} 
{\begin{tabular}{@{}ccc@{}} \toprule 
$u$  & Weight $C_{(u)}$ & Geometric Quantity $L^{(u)}_{\al\be\ga\de}$ \\ \colrule 
1  & $3 c_2+16 c_3$        & $R_{\al\be\ga\de}$ \\ 
2  & $-6-22c_2-36c_3$      & $g_{\be\de} R_{\al\ga}$ \\ 
3  & $\fr{1}{\ell^2}(6+28c_2+56c_3)$ & $g_{\al\ga}g_{\be\de}$ \\ 
4  & $-4-16c_2-32c_3$        & $e^\mu_a e_{\be b} (\altnabla_\ga e^a_\al) (\altnabla_\de e^b_\mu)$ \\ 
5  & $4+12c_2+32c_3$         & $e_{\de a} e^\mu_b (\altnabla_\al e^a_\ga) (\altnabla_\be e^b_\mu)$ \\ 
6  & $2+4c_2+8c_3$           & $g_{\be\de} e^\mu_a e^\nu_b [ (\altnabla_\al e^a_\nu) (\altnabla_\ga e^b_\mu) - (\altnabla_\ga e^a_\mu) (\altnabla_\al e^b_\nu) ]$ \\ 
\botrule 
\end{tabular}} 
\label{table1} 
\end{table} 

The action in Eq.\ \refeq{NCaction} approximately works as a model for noncommutative gravity, 
though there are some interpretational issues. 
First, it assumes that $\prt_\al\tht^{\mu\nu}=0$, 
which is a coordinate-dependent statement. 
We may try to maintain coordinate independence by requiring that $\nabla_\al\tht^{\mu\nu}=0$. 
However, such covariant-constant tensors cannot exist in most 
spacetimes.\cite{kosteleckygravity,lane} 

Second, the derivative $\altnabla$ that appears 
is covariant with respect to the SO(1,3)$_\star$ connection 
but not the Christoffel connection: 
$\altnabla_\ga{e_\al}^a = \prt_\ga{e_\al}^a + {\om_\ga}^{ab}e_{ab} 
 = {\Ga^\rh}_{\ga\al}{e_\rh}^a $. 
This means that the Christoffel symbols appear explicitly in the action. 
The troublesome terms where they appear 
violate observer-diffeomorphism symmetry, 
though they do respect {\itshape local} observer Lorentz transforms. 
For the rest of this work, 
we assume that these issues are negligible in experimentally relevant situations. 
Further, we work at quadratic order in $h_{\mu\nu}=g_{\mu\nu}-\et_{\mu\nu}$. 

To quadratic order in $h$, the gravitational SME may be written\cite{linearizedgravity} 
$S_{\rm SME} \supset \fr{1}{64\pi G_N} \int d^4x\ 
  h_{\mu\nu} \sum_{d}\widehat{\cal K}^{(d)\mu\nu\rh\si} h_{\rh\si}$.  
The noncommutative action \refeq{NCaction} contains many terms of this form,\cite{baileylane} 
though we only describe a few here: 
\begin{equation} 
S_{\rm NCR} \supset \frac{1}{64\pi G_N} \int d^4x\ h_{\mu\nu} 
 \left\{ 
 s^{(2,1)\mu\rh\nu\si} 
 + s^{(4)\mu\rh\al\nu\si\be}\prt_\al\prt_\be 
 + \cdots 
 \right\} h_{\rh\si} 
 \quad . 
\end{equation} 

First, we may match the $u=3$ mass-like term in $S_{\rm NCR}$: 
\begin{eqnarray} 
S_{{\rm NCR, mass}} 
&=& \frac{1}{64\pi G_N} \int d^4x\ 
 \left\{ 
 \left[\frac{C_{(3)}}{2\ell^6} \theta^2 \right] 
 +\left[ \frac{C_{(3)}}{2\ell^6} \left(\frac{1}{2}\tht^2\et^{\rh\si}+2{\tht_\al}^\rh\tht^{\al\si}\right) \right] h_{\rh\si} 
 \right. \nonumber \\ 
&& \left. + h_{\mu\nu} 
   \left[ s^{(2,1)\mu\rh\nu\si}+k^{(2,1)\mu\nu\rh\si} \right] 
   h_{\rh\si} 
 \right\} 
 \quad . 
\end{eqnarray} 
The first term is an irrelevant constant, 
while the 2nd corresponds to a constant stress-energy. 
The bottom line contains effective values of SME coefficients: 
\begin{eqnarray} 
s^{(2,1)\mu\rh\nu\si} &=& 
 \frac{C_{(3)}}{12\ell^4} \left[ 
 2\et^{\mu\nu}\tht^{\rh\al}{\tht^\si}_{\al} 
 +2\tht^{\rh\nu}\tht^{\si\mu} 
 +\cdots 
 \right] 
 \quad \mbox{and} \nonumber \\ 
k^{(2,1)\mu\nu\rh\si} &=& 
 \frac{C_{(3)}}{48\ell^4} \left[ 
 4\et^{\mu\nu}\tht^{\rh\al}{\tht^\si}_\al 
 +\cdots 
 \right] 
 \quad . 
\end{eqnarray} 

Second, we consider a sample kinetic effect 
with contributions from the $u=$1, 2, 4, and 5 terms: 
\begin{equation} 
S_{\rm NCR, kinetic} \supset \frac{1}{64\pi G_N} \int d^4x\ h_{\mu\nu} 
 \left\{ 
 s^{(4)\mu\rh\al\nu\si\be}\prt_\al\prt_\be 
 + \cdots 
 \right\} h_{\rh\si} 
 \quad , 
\end{equation} 
where 
\begin{equation} 
s^{(4)\mu\rh\al\nu\si\be} \sim 
 \frac{ 2C_{(1)}-3C_{(2)}+C_{(4)}+C_{(5)} }{ \ell^4 } 
 \ve^{\mu\rh\al\ka}\ve^{\nu\si\be\la} 
 \left[ 
  \tht_{\ka\ga}{\tht_\la}^\ga -\frac{1}{4}\et_{\ka\la}\tht^2 
 \right] 
 \quad . 
\end{equation} 
This coefficient regulates behavior similar to the $\bar{s}^{\mu\nu}$ coefficient 
that appears in the minimal gravitational SME.\cite{kosteleckygravity} 
We may therefore exploit existing bounds\cite{boundsbar} on $\bar{s}^{\mu\nu}$ 
to extract rough bounds on $\tht^{\mu\nu}$ 
(albeit bounds that depend on the gauge-breaking scale $\ell$): 
\begin{equation} 
\left| \frac{\tht_{\mu\nu}\tht^{\mu\nu}}{\ell^4} \right| \lsim 10^{-15} 
 \quad . 
\end{equation}

\section*{Acknowledgments}
I would like to thank Berry College and 
the IU Center for Spacetime Symmetries 
for support during the creation of this work.

\end{document}